\documentclass[10pt,conference]{IEEEtran}
\usepackage[utf8]{inputenc}
\usepackage{amsmath, amssymb}
\usepackage[pdftex]{hyperref}
\usepackage{graphicx}
\usepackage{booktabs}
\usepackage[T1]{fontenc}

\usepackage[ruled,linesnumbered]{algorithm2e}

\usepackage{xspace}
\newcommand{\rrangle}{\rangle\kern-1.2ex~\rangle\xspace}
\newcommand{\llangle}{\langle\kern-1.2ex~\langle\xspace}

\usepackage{mathtools}

\DeclarePairedDelimiterX\braket[2]{\langle}{\rangle}{#1 \delimsize\vert #2}
\DeclarePairedDelimiterX\expval[3]{\langle}{\rangle}{#1 \delimsize\vert #2 \delimsize\vert #3}

\DeclarePairedDelimiter\bbra{\llangle}{\rvert}
\DeclarePairedDelimiter\kket{\lvert}{\rrangle}
\DeclarePairedDelimiterX\bbraket[2]{\llangle}{\rrangle}{#1 \delimsize\vert #2}
\DeclarePairedDelimiterX\eexpval[3]{\llangle}{\rrangle}{#1 \delimsize\vert #2 \delimsize\vert #3}

\usepackage{color}
\definecolor{lava}{rgb}{0.81, 0.06, 0.13}

\newtheorem{theorem}{Theorem}

\SetKwComment{Comment}{}{}
\SetCommentSty{textrm}

\usepackage{ulem}

\title{Near-Minimal Gate Set Tomography Experiment Designs}
\author{\IEEEauthorblockN{Corey Ostrove, Kenneth Rudinger, Stefan Seritan, Kevin Young and Robin Blume-Kohout}
\IEEEauthorblockA{Sandia National Laboratories, Quantum Performance Laboratory\\
Albuquerque, NM 87185 and Livermore, CA 94550}}

\begin{document}

\maketitle

\begin{abstract}
Gate set tomography (GST) provides precise, self-consistent estimates of the noise channels for all of a quantum processor’s logic gates. But GST experiments are large, involving many distinct quantum circuits. This has prevented their use on systems larger than two qubits.  Here, we show how to streamline GST experiment designs by removing almost all redundancy, creating smaller and more scalable experiments without losing precision.  We do this by analyzing the “germ” subroutines at the heart of GST circuits, identifying exactly what gate set parameters they are sensitive to, and leveraging this information to remove circuits that duplicate other circuits’ sensitivities.  We apply this technique to two-qubit GST experiments, generating streamlined experiment designs that contain only slightly more circuits than the theoretical minimum bounds, but still achieve Heisenberg-like scaling in precision (as demonstrated via simulation and a theoretical analysis using Fisher information).  In practical use, the new experiment designs can match the precision of previous GST experiments with significantly fewer circuits. We discuss the prospects and feasibility of extending GST to three-qubit systems using our techniques.
\end{abstract}


\section{Introduction}

Essential to the development of the next generation of high-performance quantum computing architectures is the ability to precisely and completely characterize the operation of as-built quantum computers in the lab today. Gate set tomography (GST) is one of the most powerful tools currently available for experimentally characterizing a quantum computer, able to produce self-consistent and highly precise Heisenberg-limited estimates for all of the parameters describing a quantum computer's gate set. With this power comes experimental cost, however, which has so far limited the application of GST to characterizing systems of one or two-qubits, for example (\cite{Blume-Kohout2017-kn, Rudinger-PRX2019, xue2022quantum, mkadzik2022precision, dahlhauser2022benchmarking}). While one-qubit GST has been deployed widely in experiments on many of the main quantum computing platforms, for many hardware platforms the experimental costs associated with running two-qubit GST---which like any protocol for completely tomographing a system has resource requirements that scale exponentially in the system size---are such that it remains impractical and largely out of reach. 

In this work we introduce a technique for streamlining GST experiment designs called per-germ global fiducial pair reduction. Fiducial pair reduction (FPR) was introduced more broadly in \cite{Nielsen2020_GST} and as a class of techniques works by identifying redundancies in traditionally constructed GST experiment designs and leverages this information to prune unnecessary circuits from the design. In \cite{Nielsen2020_GST} two specific variants of FPR called per-germ and per-germ power random were introduced, and in \cite{2Qresource} their efficacy was investigated in significant detail. In the results presented in this work we find that using per-germ global fiducial pair reduction we are able to produce near-minimal experiment designs approaching information theoretic lower bounds on design size while maintaining the ability to achieve high-precision gate set estimates.

The remainder of this paper is outlined as follows. In Section \ref{sec:gst_edesign} we describe the traditional structure of GST experiment designs. In Section \ref{sec:FPR} we introduce fiducial pair reduction as a class of techniques before, in Section \ref{sec:pgg_FPR}, describing in detail the new per-germ global fiducial pair reduction method that is the focus of this work. In Section \ref{sec:results} we present results demonstrating the efficacy of experiment designs constructed using per-germ global fiducial pair reduction. In Section \ref{sec:simulations} we present end-to-end simulations of an entire GST experiment using simulated data according to different noise models, and in \ref{sec:fisher_information} present supporting theoretical analysis based on the Fisher information matrix. We close out in Section \ref{sec:discussion} with a discussion about the prospects for using per-germ global fiducial pair reduction to extend the feasible range of GST to include three-qubit systems. 

\section{Gate Set Tomography Experiment Designs}
\label{sec:gst_edesign}

\begin{figure}[t]
    \centering
    \includegraphics[width=.8\columnwidth]{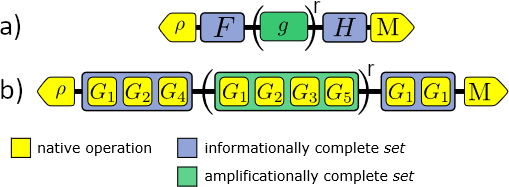}
    \caption{General structure of a gate set tomography experiment design, reproduced from Fig. 4 of \cite{Nielsen2020_GST}. $F$ denotes an informationally-complete (IC) set of state preparation fiducial circuits and $H$ denotes an IC set of measurement fiducial circuits. Each of the germs in the amplificationally-complete (AC) set $\{g\}$ is repeated $r$ times, for a number of $r$ values. In (b) we show how each of the components in $F$, $H$ and $\{g\}$ are comprised of gates from within the gate set to be characterized.}.
    \label{fig:gst_experiment_sketch_omnibus}
\end{figure}

In gate set tomography (GST) \cite{Nielsen2020_GST} we aim to construct self-consistent and highly precise estimates of all of the parameters characterizing a quantum computer's gate set $\mathcal{G} = \{ \{\kket{\rho_i} \}_{i=0}^{N_\rho-1}, \{G_k\}_{k=0}^{N_G-1}, \{\bbra{E_i^m} \}_{i,m=0}^{N_E^m, N_m}\}$. Here $\{\kket{\rho_i} \}_{i=0}^{N_\rho-1}$ denotes the set of $N_\rho$ native state preparation operations available, $\{G_k\}_{k=0}^{N_G-1}$ denotes the set of $N_G$ available gate operations, and $\{\bbra{E_i^m} \}_{i,m=0}^{N_E^m, N_m}\}$ denotes the set of $m$ native POVMs with $N_E^m$ effects for each. In practice most systems only support a single native state preparation and single native measurement (SPAM), so without loss of generality we'll make this assumption throughout the remainder of this paper.

To describe imperfect SPAM and gate operations we work in the superoperator, or transfer matrix, representation. In this representation the states are denoted using superkets, $\kket{\cdot}$, and are vectors in Hilbert-Schmidt space. Measurement effects are elements of the dual-space to states and are denoted using superbras, $\bbra{\cdot}$. Gate operations are matrices that act on elements of Hilbert-Schmidt space and evolve states through matrix multiplication. A general $d\times d$ mixed state $\rho$ corresponds to a $d^2$-dimensional vector $\kket{\rho}$ in Hilbert-Schmidt space, and as such gate operations correspond to $d^2 \times d^2$ matrices. For convenience we choose to represent our states, measurement effects and gates in the Pauli basis, i.e. the Pauli transfer matrix (PTM) representation. Given a circuit $\mathcal{C}$ consisting of some sequence of gates from $\mathcal{G}$, $\mathcal{C}= G_{l-1} G_{l-2} \cdots G_0$ (with gates listed in reverse application order), the probability of observing the $j$\textsuperscript{th} measurement outcome $\bbra{E_j}$ given initialization in the state $\kket{\rho}$ is

\begin{equation}
    p(j|\mathcal{C}) = \bbra{E_j} G_{l-1} G_{l-2} \cdots G_0 \kket{\rho}.
    \label{eqn:born_rule}
\end{equation}

As in all tomographic protocols, in order to reconstruct the gates in $\mathcal{G}$ we need an informationally-complete (IC) set of SPAM operations that span Hilbert-Schmidt space. However, very rarely in practice do actual quantum computers support an IC set of state preparations and measurements natively. As such, it is necessary to combine the available native SPAM operations with short sequences of gate operations from our gate set. The circuits used for constructing an IC set of state preparations are called \textit{preparation fiducial circuits} or \textit{prep fiducials} for short. Likewise, the circuits for constructing an IC set of measurements are called \textit{measurement fiducial circuits} or \textit{measurement fiducials}. Let $F$ be the set of prep fiducials and $H$ be the set of measurement fiducials.

In designing gate set tomography experiments our goal isn't simply to estimate the parameters of a gate set, but to do so with high precision. We achieve this by introducing a set of short circuits called \textit{germs} which are each designed to amplify our sensitivity to different subsets of the parameters of our gate set. By repeating a germ many times in a circuit,  we amplify our sensitivity to certain parameters proportionally to the number of repetitions. This allows us to estimate said parameters with Heisenberg-like precision scaling. A set of germs, $\{g\}$, that collectively amplifies our sensitivity to every amplifiable parameter of a gate set is called \textit{amplificationally complete} (AC). Let $N_a$ denote the number of amplifiable parameters in a gate set $\mathcal{G}$.

The structure of a traditional GST experiment design is shown in Figure \ref{fig:gst_experiment_sketch_omnibus}. We begin with an informationally complete set of prep and measurement fiducials, $F$ and $H$, and an amplificationally complete set of germs, $\{g\}$. We then select a set of germ powers for each germ $g_k$, $\{r_k\}$, which correspond to the number of times a germ is repeated in each circuit. Rather than setting the values of $\{r_k\}$ directly, we choose a sequence of maximum circuit depths $\{L_l\}$ (selected to be sequential powers of two) and set the germ-powers as the largest integers for which $\text{len}(g_k^{r_k}) \leq L_l$ for each $L_l$. For each repeated germ sequence we then prepend a prep fiducial $f\in F$ and append a measurement fiducial $h \in H$.

Given a complete experiment design we then experimentally measure the frequencies of each of the circuits,

\begin{equation}
    f_{jkl} \approx \eexpval{E_j^\prime}{g_k^{r_k}}{\rho},
    \label{eqn:frequencies}
\end{equation}

\noindent and use maximum likelihood estimation (MLE) to construct our estimates of the best-fit model parameters. For more on the process of fitting gate set models to experimental data see \cite{Nielsen2020_GST}. 

GST experiments give gate set estimates with Heisenberg-like precision scaling that goes like $\mathcal{O}(\frac{1}{L\sqrt{N}})$, where $N$ is the total number of shots collected per circuit and $L$ is the maximum depth used in the construction of the experiment designs. This comes, however, with nontrivial experimental cost. Setting aside the exponential scaling with system size inherent to any protocol for producing complete tomographic reconstructions of a gate set, the inclusion of multiple germ sequences and the need to tomograph these germ sequences for a number of different repetitions adds significant additional experimental overhead. In Table \ref{tab:traditional_GST_experiment_cost} we give the number of circuits required for the traditionally constructed two-qubit GST experiments used later on in this work. The high resource requirements associated with traditional GST experiments has so far limited its experimental deployment to characterizing subsystems of at most two qubits, and even then two-qubit GST still remains too onerous on certain quantum computing hardware platforms. 

In the remainder of this work we explore a new technique for reducing the experimental overhead for running GST. This technique, called \textit{per-germ global fiducial pair reduction}, identifies and prunes redundant or unnecessary circuits from a traditionally constructed GST experiment design, significantly decreasing the number of circuits required while maintaining high precision.

\begin{table}[htbp]
  \centering
  \caption{Experimental cost of two-qubit GST using XYCPHASE gate set with $L=64$.}
    \begin{tabular}{p{.4\columnwidth}c}
    \toprule
    Experiment Design & \multicolumn{1}{l}{Number of Circuits} \\
    \midrule
    Robust Germ Set w/o FPR & 31290 \\[.5em]
    Standard Germ Set w/o FPR & 7722 \\[.5em]
    Standard Germ Set w/ Per-Germ Global FPR & 2576 \\
    \bottomrule
    \end{tabular}%
  \label{tab:traditional_GST_experiment_cost}%
\end{table}%

\section{Streamlining Experiments with Fiducial Pair Reduction}

As mentioned in the previous section, GST experiments, as traditionally constructed, come with large experimental overhead. In this section we'll describe the theory behind a technique for streamlining GST experiment designs called fiducial pair reduction (FPR). In Section \ref{sec:FPR} we introduce the general theory behind FPR and in Section \ref{sec:pgg_FPR} we describe the theory and algorithmic implementation of a new variant of FPR called per-germ global FPR, which is the main subject of this work. 

\subsection{Fiducial Pair Reduction}
\label{sec:FPR}

Fiducial pair reduction  is based on the following pair of observations about the structure of GST experiment designs:

\begin{enumerate}
    \item In GST experiments we select sets of preparation fiducials and measurement fiducials, $F$ and $H$, such that they are both informationally complete. Each repeated germ sequence is then sandwiched between every pair of elements from $F$ and $H$. That is, for each repeated germ sequence we have sufficient information to perform quantum process tomography (QPT) on that sequence.
    
    \item The germs used in GST experiments are specifically designed to amplify sensitivity to particular  subsets of gate set parameters, and the number of parameters amplified by any given germ is typically relatively small compared to the number of parameters in the gate set overall. 
\end{enumerate}

\noindent Taken together, these observations point to the conclusion that for any given germ sequence most of the pairs of prep and measurement fiducials provide relatively little value, as they probe parameters of our gate set that that germ is not sensitive to. FPR leverages this fact by identifying the parameters of our gate set each germ is sensitive to and then selecting for each germ a subset of the full set of prep and measurement fiducials that act as probes for those particular parameters.  In doing so we can achieve significant experimental savings without sacrificing our ability to generate high-precision estimates of the gate set parameters. In the next section we dive deeper into the theory behind FPR and will describe a particular implementation called per-germ global FPR that achieves additional streamlining by identifying redundancies in the construction of the germ set itself.

\subsection{Per-Germ Global FPR}
\label{sec:pgg_FPR}

FPR leverages the fact that any given germ is only sensitive to relatively few of the parameters of a gate set.  In order to determine which fiducial pairs we want to keep for a given germ, we must therefore be able to answer the question.  \textit{``Given a germ, $g$, which parameters does that germ amplify sensitivity to?''} We answer this question in what follows below.

Let $\tau(g)$ be the Pauli transfer matrix representation of $g$. Consider the Jacobian of $\tau(g)^r$ with respect to the model parameters (for some germ-power $r$), which we define as

\begin{equation}
    \vec{\nabla}_g^{(r)} \equiv \frac{1}{r} \frac{\partial (\tau(g)^r)}{\partial \vec{\theta}},
    \label{eqn:germ_jacobian}
\end{equation}

\noindent where $\vec{\theta}$ is a vector of gate set parameters of length $N_p$, and the derivatives are evaluated at the target model, $\vec{\theta}_0$. The inclusion of the normalization factor of $\frac{1}{r}$ is not standard (or necessary), but makes a connection to group theory more clear shortly. This Jacobian is a tensor-valued quantity, and we can consider a single slice through this tensor for the derivative with respect to a particular parameter $\theta_n$, $\frac{1}{r}\frac{\partial \tau(g)^r}{\partial \theta_n}$. This can be rewritten using the product rule as

\begin{equation}
    \frac{1}{r}\frac{\partial (\tau(g)^r)}{\partial \theta_n} = \frac{1}{r} \sum_{i=0}^{r-1} \left( (\tau(g))^i \frac{\partial \tau(g)}{\partial \theta_n} (\tau(g))^{-i} \right) \tau(g)^{r-1}.
    \label{eqn:tensor_slice_group_avg}
\end{equation}

\noindent As $r\rightarrow \infty$ the set $\{\tau(g)^r\}$, forms an approximate representation of the cyclic group (recall that $\tau(g)$ is a representation of a unitary). Thus in the $r\rightarrow \infty$ limit Equation \ref{eqn:tensor_slice_group_avg} corresponds to a group average of $\frac{\partial \tau(g)}{\partial \theta_n}$ (up to an unimportant change of basis from the final multiplication by $\tau(g)^{r-1}$). Applying Schur's lemma, this group average projects $\frac{\partial \tau(g)}{\partial \theta_n}$ onto the commutant of $\tau(g)$, the subalgebra of all of the matrices which commute with $\tau(g)$ \cite{gambetta2012simultaneousRB}. This allows us to connect the asymptotic sensitivity of a germ to parameters in our gate set to its sensitivity for $r=1$ through the projection of the Jacobian for that germ onto its commutant. 

Let $\vec{\nabla}_g^{\infty}$ be the projection of each of the slices of $\vec{\nabla}_g^{(1)}$ onto the commutant of $g$. We can matricize $\vec{\nabla}_g^{\infty}$ to convert it from a $d^2\times d^2 \times N_p$ tensor to a $d^4\times  N_p$ matrix (where unambiguous we will use $\vec{\nabla}_g^{\infty}$ to denote both the tensor and equivalent matrix interchangeably). We call $\vec{\nabla}_g^{\infty}$ the \textit{twirled derivative} of the germ $g$. The right singular vectors of $\vec{\nabla}_g^{\infty}$ correspond to directions in parameter space asymptotically amplified by $g$ and the singular values indicate the amount of amplification.

Using the twirled derivative we can identify the subset of gate set parameters (or linear combinations thereof) that a germ amplifies---when referring to a germ's amplified parameters we typically mean the linear combinations of gate set parameters it is sensitive to. This is used in the germ selection process in order to construct a set of amplificationally complete germs, but will also be essential in performing FPR. 

For FPR we assume that we've already previously performed germ selection and are given as input an AC set of germs, $\{g\}$. Our goal is to identify, for each germ, a reduced set of fiducial pairs which gives us sensitivity to each of that germ's amplified parameters. Before doing so, however, there is actually one more source of redundancy in GST experiment designs that the ability to identify a germ's amplified parameters allows us to reduce, and it is here that the per-germ global FPR technique diverges from the techniques introduced in \cite{Nielsen2020_GST, 2Qresource}. That is the redundancy within the germ set itself. In the construction of amplificationally complete germ sets it is essentially never the case that every germ amplifies a wholly disjoint set of parameters from all of the other germs. It is generically the case that multiple germs will partially overlap in the parameters that they amplify. Were we to construct sets of fiducial pairs that guaranteed sensitivity to \textit{all} of a germ's amplified, as in \cite{Nielsen2020_GST, 2Qresource}, then overlap with another germ's amplified parameters inevitably would introduce some amount of redundancy into our experiment design. In per-germ global FPR we therefore streamline experiment designs using a two-stage process. In the first stage we remove the redundancy within a germ set by identifying for each germ a subset of its amplified parameters disjoint (i.e. linearly independent) from all of the others such that collectively these subsets still span the space of amplifiable gate set parameters. In the second stage we identify reduced sets of fiducial pairs for each germ while requiring only sensitivity to the subset of that germ's amplified parameters identified as necessary in the first stage.

We begin the first stage of per-germ global FPR by calculating the twirled derivatives of each germ in our germ set $\{g\}$, $\{\vec{\nabla}^{\infty}_g\}_{g \in \{g\}}$. To get the amplified parameters for each germ we then take the compact SVD of each of these twirled derivatives in order to obtain their (non-trivial) right singular vectors. Let $V_g$ be the matrix whose columns correspond to the (non-trivial) right singular vectors of $\vec{\nabla}^{\infty}_g$. Let $J$ be the matrix formed by concatenating each $V_g$ column-wise, i.e.

\begin{equation}
    J = \begin{bmatrix}V_{g_0} & V_{g_1} & \cdots & V_{g_{N_g}} \end{bmatrix},   
    \label{eqn:sing_vector_concat_matrix}
\end{equation}

\noindent where $N_g$ is the number germs in the germ set. The problem of identifying a disjoint set of amplified parameters for each germ such that they collectively span the space of amplifiable gate set parameters is equivalent to the problem of selecting a subset of $N_a$ columns from $J$ such that the resulting submatrix has full column rank and is well-conditioned. This is a version of the well-known column subset selection problem (CSSP), widely studied in computer science. While formally NP-complete \cite{shitov2021column}, CSSP is a problem that naturally arises in a many domains ranging from numerical linear algebra, to machine learning, to optimal experiment design \cite{ ordozgoiti2019regularized,de2011note,avron2013faster}. As such, there are a number of well-studied heuristics for tackling this problem. In this work we use two such heuristics, one a greedy search heuristic using low-rank updates, and the other a heuristic based on rank-revealing QR decompositions. For details on these CSSP heuristics see Appendix \ref{app:cssp_heuristics}. Once we have identified a subset of the columns of $J$ satisfying the CSSP we've constructed we simply associate these columns back to their corresponding germs. The details of this algorithm are summarized in Algorithm \ref{alg:pgg_fpr_stage_1}.

\begin{algorithm}
\caption{Per-Germ Global FPR Stage One}\label{alg:pgg_fpr_stage_1}
\SetKwInOut{Input}{Input}\SetKwInOut{Output}{Output}
\Input{Gate set $\mathcal{G}$, an AC set of germs $\{g\}$ for $\mathcal{G}$.}
\Output{\{$W_g$\}, subsets of amplified parameters for each $g$ such that $\bigcup_{g} W_g$ spans the space of amplifiable parameters.}
$J \gets [\hspace{.25em}]$\;
\For{$g\in \{g\}$}{
    $\vec{\nabla}^{\infty}_g \gets$ \texttt{compute\_twirled\_deriv}$(g)$ \Comment*{Eqn.~\ref{eqn:tensor_slice_group_avg}}
    $V_g \gets $ \texttt{compute\_right\_sing\_vecs} $(\vec{\nabla}^{\infty}_g)$ \Comment*{Using SVD}
    $J \gets $ \texttt{concatenate}$(J,V_g)$ \Comment*{Column-wise}
    }
\If{search\_mode $=$ ``greedy''}{
$J_{min} \gets $ \texttt{greedy\_search}$(J)$ \Comment*[r]{See Appendix \ref{app:greedy_search}}
}
\ElseIf{search\_mode $=$ ``rank-revealing-QR''}{
$J_{min} \gets $ \texttt{rrqr\_css}$(J)$ \Comment*{See Appendix \ref{app:rrqr}}
}
\For{$g\in \{g\}$}{
    $W_g \gets$ \texttt{map\_columns\_to\_germs}$(J_{min},g)$ \Comment*{Map parameters in $J_{min}$ to associated germs.}
    }
\end{algorithm}

In the second stage of per-germ global FPR we utilize the reduced sets of amplified parameters for each germ that we constructed in the first stage and construct reduced sets of fiducial pairs for each germ with sensitivity to these particular parameters. We begin with informationally complete sets of preparation and measurement fiducials, $F$ and $H$. In order for a fiducial pair to be sensitive to an amplified parameter for a germ it must be the case that the output probability distribution for the circuit corresponding to that germ-fiducial pair combination must have linear sensitivity to changes in that amplified parameter. That is, the directional derivative for the circuit's output probability distribution in the direction in parameter space corresponding to that amplified parameter must be non-trivial. To construct a complete set of fiducial pairs for a germ $g$ we then proceed as follows. Start by constructing the set of circuits corresponding to sandwiching $g$ between each element of $F$ and $H$, $\mathcal{C}_g = \left\{ h g f \right\}_{f\in F, h \in H}$, and then calculate the output probability distribution for each $c\in \mathcal{C}_g$:

\begin{equation}
    \vec{p}_{g,c} = \begin{bmatrix} \bbra{E_0} \\ \bbra{E_1} \\ \vdots \\ \bbra{E_{N_E-1}}  \end{bmatrix} c \kket{\rho}.
    \label{eqn:prob_dist_fid_pair}
\end{equation}

\noindent Next we calculate the Jacobians of each of these probability distributions $\vec{p}_{g,c}$ with respect to the vector of gate set model parameters $\vec{\theta}$ evaluated at the target model $\vec{\theta}_0$, 

\begin{equation}
    J_{g,c} = \frac{\partial \vec{p}_{g,c}}{ \partial \vec{\theta}}.
    \label{eqn:fid_pair_jacobian}
\end{equation}

\noindent Let $W_g$ be the matrix whose columns correspond to the subset of the right singular vectors for $g$'s twirled derivative that we selected in stage one. To evaluate the sensitivity of each fiducial pair to these parameters we calculate directional derivatives of the probability distribution in those directions,

\begin{equation}
    D_{g,c} = J_{g,c} W_g.
    \label{eqn:directional_derivatives}
\end{equation}

\noindent To evaluate the sensitivity for a set of of multiple fiducials we can construct a composite Jacobian by concatenating the Jacobians for each of the constituent fiducial pairs row-wise,

\begin{equation}
    D_{g,\{c\}} = \begin{bmatrix} D_{g,c_1} \\ D_{g,c_2} \\ \vdots \end{bmatrix}.
    \label{eqn:fid_set_composite_jacobian}
\end{equation}

\noindent Our goal then is to identify a set of circuits (i.e. fiducial pairs) $\{c\} \subset \mathcal{C}_g$ such that $D_{\{c\}}$ has full column rank and is well-conditioned. As before, constructing such a $\{c\}$ optimally while minimizing the size of the fiducial pair set is a challenging combinatorial optimization problem. As with the CSSP we encountered in stage one we use a greedy search heuristic combined with low rank update methods to construct a set of fiducial pairs satisfying the requisite constraints; see Appendix \ref{app:cssp_heuristics} for more details. The details of stage two are summarized below in Algorithm \ref{alg:pgg_fpr_stage_2}.

\begin{algorithm}
\caption{Per-Germ Global FPR Stage Two}\label{alg:pgg_fpr_stage_2}
\SetKwInOut{Input}{Input}\SetKwInOut{Output}{Output}
\Input{Gate set $\mathcal{G}$, an AC set of germs $\{g\}$, subsets of amplified parameters $\{W_g\}$ for each germ, IC prep fiducials $F$ and measurement fiducials $H$.}
\Output{$\{(f_l, h_k)\}_{g\in\{g\}}$, where $\{(f_l, h_k)\} \subseteq F\times H$ are reduced sets of fiducial pairs for the germ $g$ that give sensitivity to every parameter in $W_g$.}

\For{$g \in \{g\}$}{
    $\mathcal{C}_g \gets \{\}$\;
    $D_{\text{cand}} \gets \{\}$ \Comment*{Candidate directional derivative matrices}
    \For{$(f_i, h_j) \in F\times H$}{
        $c \gets h_j g f_i$ \Comment*{Circuit for fiducial pair}
        $\mathcal{C}_g \gets$ \texttt{append}$(\mathcal{C}_g , f_i g h_j)$ \Comment*{Add $c$ to set $\mathcal{C}_g$}
        $J_{g,c} \gets$ \texttt{compute\_jacobian}$(\vec{p}_{g,c})$ \Comment*{Eqn. \ref{eqn:fid_pair_jacobian}}
        $D_{\text{cand}} \gets$ \texttt{append}$(D_{\text{cand}}, J_{g,c} W_g)$ \Comment*{Add to set of candidate directional derivatives}
    }
    $D_{g, \{c\}}, \{(f_l, h_k)\} \gets$ \texttt{greedy\_search}$(D_{\text{cand}}, \mathcal{C}_g)$ \Comment*{Use greedy algorithm to select subset of fiducial pairs, see Appendix \ref{app:greedy_search}.}
}
\end{algorithm}

\section{Results}
\label{sec:results}

In this section we will present results demonstrating the efficacy of GST experiment designs constructed using per-germ global FPR and compare their performance to traditionally constructed experiment designs. In Section \ref{sec:simulations} we present results from end-to-end runs of GST using data from simulated noise models. In \ref{sec:fisher_information} we analyze the performance of these reduced experiment designs theoretically using an analysis of the Fisher information. In both cases we'll show that the experiment designs constructed using per-germ global FPR can generate high precision estimates of our gate set parameters while significantly reducing the experimental cost of GST.

\subsection{Simulations}
\label{sec:simulations}

In this section we demonstrate the efficacy of various GST experiment designs by performing end-to-end simulations of the GST fitting process using data simulated under a number of different noise models. For the two-qubit simulations we use a gate set consisting of serialized single-qubit $R_x(\pi/2)$ and $R_y(\pi/2)$ rotations on each of the individual qubits and a two-qubit CPHASE gate, which we'll refer to as the XYCPHASE gate set. 

We simulate GST under two different classes of noise model. The first is a coherent-only noisy model in which each of the $15$ two-qubit Hamiltonian error generators (see \cite{rbk2022taxonomy} for more on the error generator representation) is sampled independently according to a normal distribution with zero-mean and a standard deviation of $.01$. The second includes both coherent and stochastic errors. For this model we sample the coherent error rates just as for the coherent-only model, but we additionally sample a rate for each of the $15$ stochastic error generators uniformly at random from the interval $[0, 10^{-4}]$ (while checking to ensure complete-positivity). For the simulations below we have sampled six random instantiations for each of these classes of noise model, and for each of these randomly sampled noise models collected a set of simulated data to fit against using GST.

We consider three different experiment designs, each with an $L$ of 64. The first two experiment designs are examples of traditionally constructed GST experiment designs without the application of FPR. The difference between these two is in the choice of the germ set. For one of these two experiment designs we use a set of 15 germs that we refer to as the standard germ set, while for the latter we use a much larger set of 68 germs that has stronger robustness guarantees (but is significantly more expensive). For more on the definitions and constructions of these germ sets see \cite{Nielsen2020_GST}. For the final experiment design we apply per-germ global FPR to the design using the standard germ set. Table \ref{tab:traditional_GST_experiment_cost} summarizes the total number of circuits used for each of these experiment designs. Noteworthy is the fact that the experiment designs constructed using per-germ global FPR approach the information theoretic lower bound for the number of circuits required. Each additional iteration (i.e. each $L$) of the per-germ global FPR experiment design adds a total of 394 circuits. The information theoretic lower bound is $\frac{N_a}{N_E-1}$, where for the XYCPHASE gate set $N_a=1023$ and $N_E=4$, giving a lower bound of 341 circuits per iterations. As such, with the per-germ global FPR experiment design we are within just 53 circuits (per-iteration) of the smallest possible GST experiment for this gate set.

\begin{figure}
    \centering
    \includegraphics[width= .75 \columnwidth]{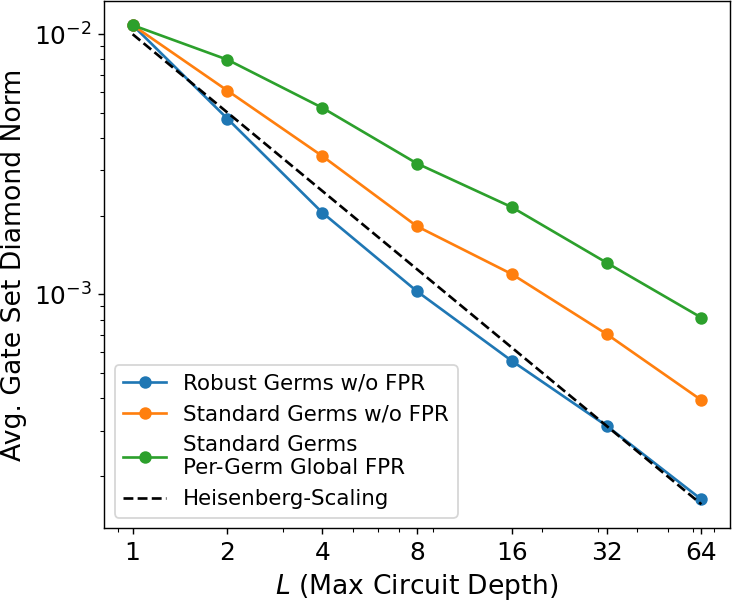}
    \caption{Average diamond distance between estimated and true data generating gate set versus $L$ for the coherent-only noise models. Included are two variants of traditionally constructed GST experiment designs using two different variants on the germ set (standard and robust, see main text) and an experiment design constructed using per-germ global FPR applied to the standard germ set. A line indicating Heisenberg-like $\sim \frac{1}{L}$ scaling is added as a guide. All three experiment designs achieve the expected asymptotic Heisenberg-precision scaling.}
    \label{fig:2Q_avg_ddists_vs_L_H}
\end{figure}

For a well-constructed GST experiment designs we expect that, in the absence of stochastic errors, the precision of our estimates should increase linearly in the maximum depth of the experiment design. In Figure \ref{fig:2Q_avg_ddists_vs_L_H} we plot the average gate set diamond distance between our estimates and the truth for the coherent-only noise model as a function of $L$, averaged over the six random instantiations for the noise model. Average gate set diamond distance is defined as

\begin{equation}
    \| \hat{\mathcal{G}} - \tilde{\mathcal{G}} \|_{\diamond} \equiv \frac{1}{N_G} \sum_{i=0}^{N_G-1} \| \hat{G}_i - \tilde{G}_i \|_{\diamond},
\end{equation}

\noindent where $\hat{\mathcal{G}}$ and $\tilde{\mathcal{G}}$ are the estimated and true data generating gate sets respectively, $\hat{G}_i$ and $\tilde{G}_i$ are their respective constituent gates, and $\| \cdot \|_\diamond = \max\limits_{X\;\| X\|_1 \leq 1} \| (\cdot \otimes I_n) X \|_1$ is the standard definition of the diamond norm \cite{aharonov1998quantum}\footnote{Some care must be taken when using diamond distance to between two gate sets as a measure of precision, as this is a generally gauge-variant quantity. In reporting diamond distances here we are using a gauge fixing technique called gauge-optimization. See \cite{Nielsen2020_GST} for more on gauge freedoms for gate sets.}.

\begin{figure}
    \centering
    \includegraphics[width=.75\columnwidth]{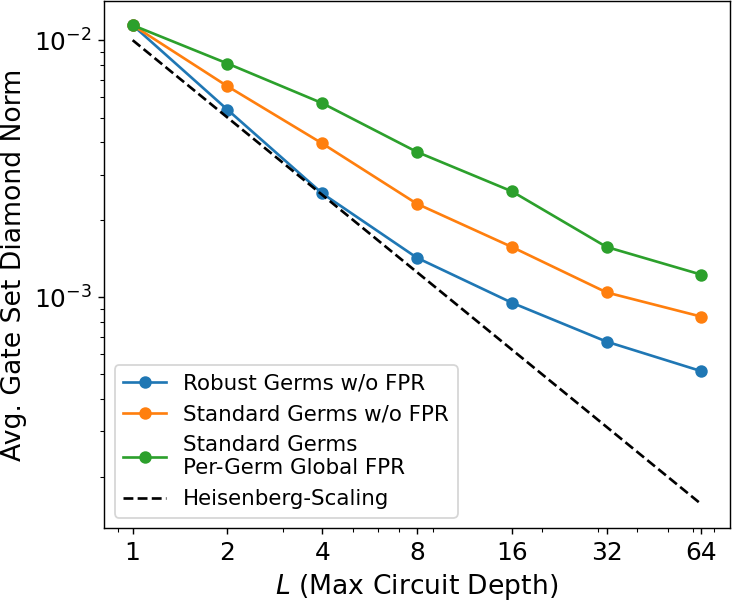}
    \caption{Average diamond distance between estimated and true data generating gate set versus $L$ for the noise models with both coherent and stochastic noise. We plot here the same experiment designs as in Figure \ref{fig:2Q_avg_ddists_vs_L_H}. Unlike the coherent-only error model, in the presence of stochastic noise we no longer expect Heisenberg-like precision scaling indefinitely, but rather (as seen here) for the precision of our estimates to plateau at long circuit depths as we become decoherence limited by the stochastic noise rates.}
    \label{fig:2Q_avg_ddists_vs_L_HS}
\end{figure}

From Figure \ref{fig:2Q_avg_ddists_vs_L_H} we can see that all three experiment designs indeed achieve the desired $\mathcal{O}(\frac{1}{L})$ asymptotic scaling of precision with circuit depth desired for achieving Heisenberg-like scaling. In Figure \ref{fig:2Q_avg_ddists_vs_L_HS} we plot the average gate set diamond distance between our estimates and the truth (averaged over the instantiations of the noise model) for the noise model including both coherent and stochastic errors. Unlike when we're subject to only coherent noise, here we no longer expect to achieve Heisenberg-like precision scaling indefinitely, as the presence of stochastic noise
imposes a decoherence limit such that for sufficiently long circuits we eventually lose all resolution on certain parameters of a gate set. As such, with stochastic noise included we expect a well-constructed GST experiment design to initially increase its precision linearly with $L$ before beginning to plateau as we approach the decoherence limit, exactly as seen in Figure \ref{fig:2Q_avg_ddists_vs_L_HS}.

While achieving a certain asymptotic precision scaling for our experiment designs is something of theoretical interest, in practice it is often the case that the primary constraint on experimental resources for a quantum characterization experiment isn't the maximum depth of the circuits used in an experiment design, but rather the total number of circuits used overall. In this case the primary goal in designing a GST experiment design is instead to allow an experimentalist to achieve the most bang for their buck possible given a particular circuit budget. Here per-germ global FPR is the clear winner, as explained below. 

\begin{figure}
    \centering
    \includegraphics[width = .75 \columnwidth]{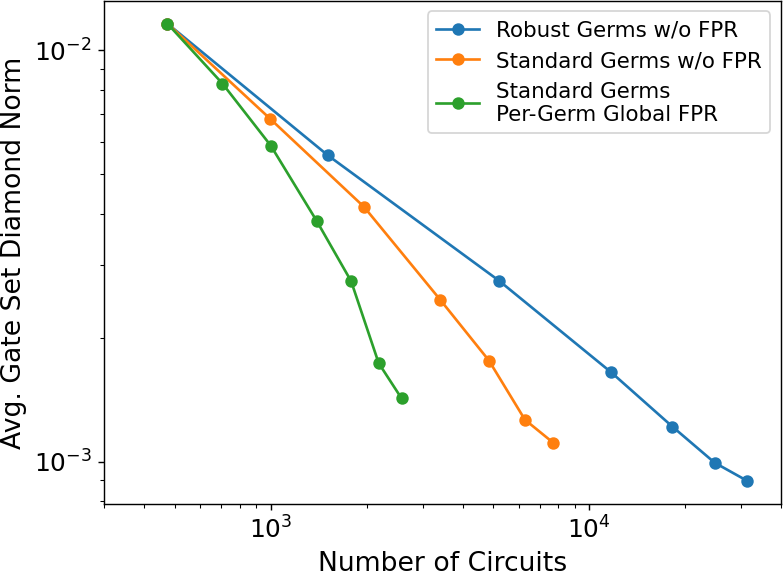}
    \caption{Average diamond distance between estimated and true data generating gate set versus circuit count for the coherent-only noise models. The experiment design constructed using per-germ global FPR achieves a final precision comparable to two traditionally constructed experiment designs using $\sim 2.4$ and $\sim 7.1$ times fewer circuits than the standard and robust germ sets without FPR applied.} 
    \label{fig:2Q_avg_ddists_vs_num_ckts_H}
\end{figure}

\begin{figure}
    \centering
    \includegraphics[width = .75 \columnwidth]{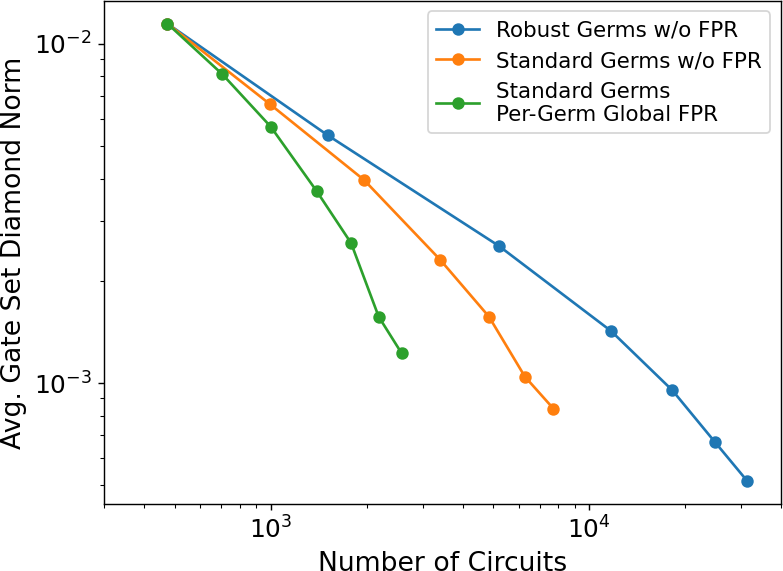}
    \caption{Average diamond distance between estimated and true data generating gate set versus circuit count for the noise models with both coherent and stochastic noise. The experimental savings of the per-germ global FPR design is comparable to what was found in Figure \ref{fig:2Q_avg_ddists_vs_num_ckts_H} for the coherent-only noise model.}
    \label{fig:2Q_avg_ddists_vs_num_ckts_HS}
\end{figure}

In Figures \ref{fig:2Q_avg_ddists_vs_num_ckts_H} and \ref{fig:2Q_avg_ddists_vs_num_ckts_HS} we re-plot the average gate set diamond distance to the truth from Figures \ref{fig:2Q_avg_ddists_vs_L_H} and \ref{fig:2Q_avg_ddists_vs_L_HS}---for the coherent-only and coherent plus stochastic noise models respectively---instead as a function of the total number of circuits used overall in the experiment design. Each point along the plotted curves corresponds to a particular $L$ value from Figures \ref{fig:2Q_avg_ddists_vs_L_H} and \ref{fig:2Q_avg_ddists_vs_L_HS}, with $L$ increasing from from left-to-right. Comparing these three experiment designs in Figure \ref{fig:2Q_avg_ddists_vs_L_H} we find that the experiment design using per-germ global FPR achieves a final precision at $L=64$ comparable to that achieved by the traditionally constructed experiment designs at $L=32$ and $L=16$ for the standard and robust germ sets respectively. However, at $L=64$ the per-germ global FPR design uses a total of $2576$ circuits, while the traditional standard germ set design at $L=32$ uses $6282$ circuits and the robust germ set at $L=16$ uses $18234$. As such, the experiment design using per-germ global FPR uses $\sim 2.4$ and $\sim 7.1$ times fewer circuits respectively, while achieving comparable levels of precision\footnote{For practical reasons related to the MLE we typically do not apply FPR to the $L=1$ circuits in an experiment design, so we actually anticipate these savings to increase slightly as we go out to larger circuit depths and those costs are amortized.}. Figure \ref{fig:2Q_avg_ddists_vs_num_ckts_HS} tells largely the same story for the noise model with coherent and stochastic errors.

\subsection{Fisher Information Analysis}
\label{sec:fisher_information}

The Fisher information matrix is a widely used tool for analyzing the performance of experiment designs \cite{ly2017tutorial}. Abstracting away the structure, a GST experiment consists of a set of circuits $\mathcal{C}$ where for each circuit $c \in \mathcal{C}$ we collect $N_c$ clicks, giving a vector $\vec{n}_c$ aggregated by $N_E$ different possible measurement outcomes. Let $\mathcal{L}(\vec{n}_c|\vec{\theta})$ be the likelihood that a gate set with parameters $\vec{\theta}$ generated the observed data for $c$. The likelihood $\mathcal{L}(\vec{n}_c|\vec{\theta})$ is the probability of sampling a set of counts $\vec{n}_c$ from a multinomial distribution with outcome probabilities $p_c(\vec{\theta})$,

\begin{equation}
    \mathcal{L} (\vec{n}_c|\vec{\theta})= \frac{N_c!}{\vec{n}_c!} \prod_{i=0}^{N_E-1} p_{c,i}^{n_{c,i}}(\vec{\theta}),
    \label{eqn:ckt_likelihood}
\end{equation}

\noindent where $\vec{n}_c!= n_{c,0}! n_{c,1}! \cdots n_{c,N_E-1}!$, and where $p_{c,i}$\footnote{Where unambiguous we'll suppress the functional dependence of $p_{c,i}$ on $\vec{\theta}$.} and $n_{c,i}$ are the predicted probability and empirical count for the $i$\textsuperscript{th} circuit outcome. Instead of directly working with the likelihood, it is standard for convenience to instead work in terms of the log-likelihood $l(\vec{n}_c|\vec{\theta})$,

\begin{equation}
l(\vec{n}_c|\vec{\theta}) = \log(N_c!) - \sum_{i=0}^{N_E-1} \log(n_{c,i}!) +  \sum_{i=0}^{N_E-1} n_{c,i} \log(p_{c,i}(\vec{\theta})).
\label{eqn:ckt_log_likelihood}
\end{equation}

The Fisher information matrix measures the sensitivity of an experiment design to changes in the parameters of a model at some point in parameter space, and (under suitable regularity conditions) is proportional to the expectation value over the observed data of the Hessian of the log-likelihood function,

\begin{equation}
     (I_c)_{i,j}(\vec{\theta}) = -\mathbb{E}_{\vec{n}_c}\left[\frac{\partial^2}{\partial \theta_i \partial \theta_j}  l(\vec{n}_c|\vec{\theta}) \right].
     \label{eqn:fisher_info_def_hessian}
\end{equation}

\noindent Plugging the log-likelihood from Equation \ref{eqn:ckt_log_likelihood} into Equation \ref{eqn:fisher_info_def_hessian} gives the Fisher information matrix for the circuit $c$,

\begin{equation}
    I_c= N_c \sum_{i=0}^{N_E-1} \left( \frac{1}{p_{c,i}} \left(\frac{\partial p_{c,i}}{\partial \vec{\theta}}\right) \left(\frac{\partial p_{c,i}}{\partial \vec{\theta}}\right)^\top - H_i \right),
    \label{eqn:per_ckt_fisher_info}
\end{equation}

\noindent where $H_i$ is the hessian of $p_{c,i}$ with respect to $\vec{\theta}$. The Fisher information matrix for an entire experiment is simply the sum over those for each circuit,

\begin{equation}
    I_{\mathcal{C}}=\sum_{c \in \mathcal{C}} I_c.
    \label{eqn:complete_experiment_fisher_info}
\end{equation}

The power of the Fisher information matrix in analyzing experiment designs follows from the Cram\'er-Rao bound, which states

\begin{equation}
    \Sigma \geq I_{\mathcal{C}}^{-1},
    \label{eqn:cramer-rao}
\end{equation}

\noindent where $\Sigma$ is the covariance in our estimate of $\vec{\theta}$, and the matrix inequality means that $\Sigma-I_\mathcal{C}^{-1}$ is positive semi-definite. In other words, the inverse of the Fisher information matrix places a lower bound on the achievable precision in our estimation of a gate set's parameters.

The eigenvectors and eigenvalues of $\Sigma$ give the axes and size of the uncertainty ellipsoid for our estimate, which implies by the Cram\'er-Rao bound that for a well-constructed GST experiment design we should find that the eigenvalues of the Fisher information matrix that correspond to amplifiable parameters increase linearly with the maximum circuit depth of the experiment design for Heisenberg-like precision scaling. 

\begin{figure}
    \centering
    \includegraphics[width=.75 \columnwidth]{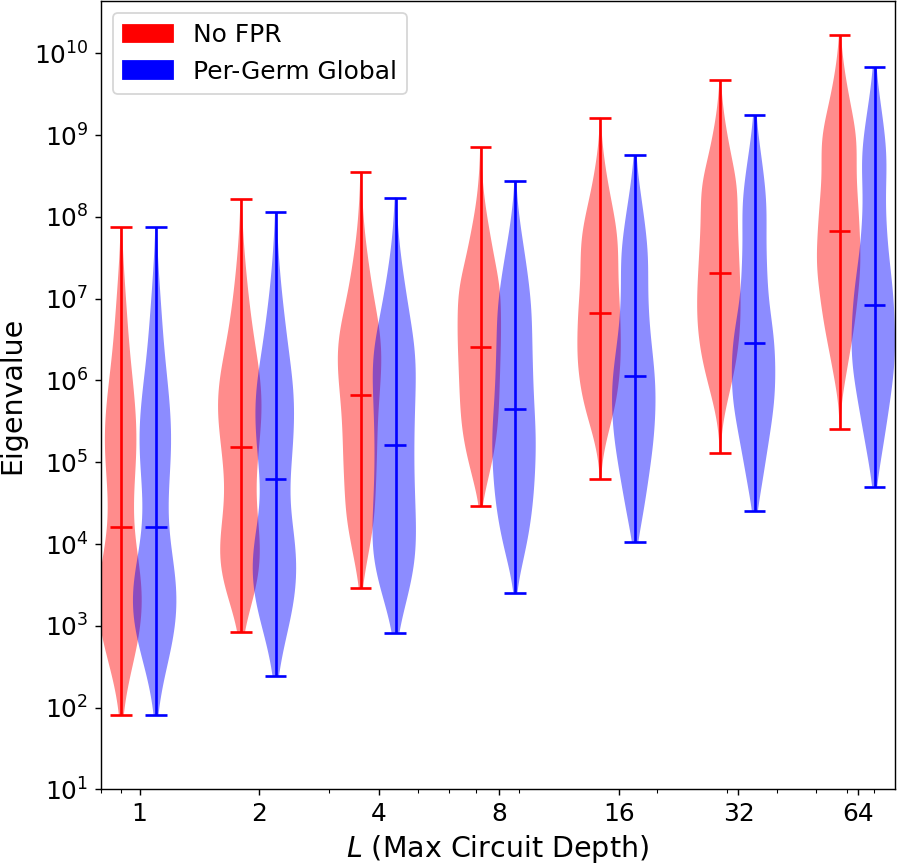}
    \caption{Comparison of the spectra of the Fisher information matrices for experiment designs without FPR applied and with per-germ global FPR applied for the standard germ set. For a well-constructed experiment design we expect that the eigenvalues of the Fisher information matrix associated with amplifiable parameters should increase linearly when increasing $L$ for the experiment. For both the experiment designs without FPR and with per-germ global FPR we do see a linear increase in the spectra of the Fisher information as expected.}
    \label{fig:2Q_fisher_info_spectra_dist_comparision}
\end{figure}

We analyze here the Fisher information matrices for the same two-qubit experiment designs we simulated in Section \ref{sec:simulations}. Due to the very large number of parameters ($\sim 1000$) for the two-qubit XYCPHASE gate set (and for two-qubit gate sets more generally) it isn't possible to usefully plot the eigenvalues of the Fisher information directly. In Figure \ref{fig:2Q_fisher_info_spectra_dist_comparision} we instead plot the distribution of the eigenvalues of the Fisher information matrix for the experiment design using the standard germ set without FPR and for the design using per-germ global FPR as a function of $L$. We additionally include points corresponding to the median and the extremal points of the distributions. From Figure \ref{fig:2Q_fisher_info_spectra_dist_comparision} we see that for both experiment design the distribution of the Fisher information eigenvalues indeed increases linearly with circuit depth as we expected. Worth noting is the apparent offset between the designs with and without FPR applied. This is primarily due to the fact that the designs without FPR applied simply contain significantly more data overall (which factors into the Fisher information through Equation \ref{eqn:per_ckt_fisher_info}). Were we to normalize the total number of counts overall between the two experiments we expect this offset would essentially disappear.

\section{Discussion}
\label{sec:discussion}

To close out this paper we'll comment now briefly on the prospects of scaling up GST to the characterization of three-qubit systems using per-germ global FPR. While moving from the complete characterization of two-qubit to three-qubit systems sounds like a small step, the exponential scaling the the resource requirements of GST with system size mean that in practice three-qubit GST is significantly more challenging. Consider, for example, a three qubit gate set consisting of (serialized) single-qubit $R_x(\pi/2)$ and $R_y(\pi/2)$ rotations on each qubit together with the three-qubit fCSWAP gate \cite{warren2023extensive}. We'll call this gate set XYfCSWAP for short. Whereas the two-qubit XYCPHASE gate set studied in this work has $\sim 1000$ parameters, the (superficially similar sounding) XYfCSWAP gate set has $\sim 29000$ parameters. Necessarily then the experimental resources required to characterize the three-qubit XYfCSWAP gate set are likewise significantly greater than for two-qubit gate sets. It is nonetheless still possible to construct traditional GST experiment designs for these systems, and for the XYfCSWAP gate set have found such an experiment design using the standard germ set requiring $69620$ circuits. Applying per-germ global fiducial pair reduction to the experiment design for the XYfCSWAP gate, however, results in an experiment design containing $23272$ circuits, a factor of $\sim 3$ times fewer. While this is clearly far out of reach for some platforms, many platforms based on superconducting qubits or semiconductor qubits have experimental throughputs sufficiently high that an experiment of this size is well within their practical capabilities (and such platforms in fact often perform characterization experiments that are substantially more expensive than even this \cite{hashim2022benchmarking, xue2022quantum}). It is perhaps convenient then that on many of these same superconducting qubit based platforms the implementation of high-quality native three-qubit gate operations is now under active investigation \cite{warren2023extensive, kim2022high}, and as such are well-positioned to take advantage of the unique capabilities GST provides for characterizing and improving the operation of these gates. While such an experimental demonstration is out-of-scope for this work, we look forward to producing such a demonstration in future work.


\section{Acknowledgements}
This research was funded, in part, by the Office of the Director of National Intelligence (ODNI), Intelligence Advanced Research Projects Activity (IARPA). All statements of fact, opinion or conclusions contained herein are those of the authors and should not be construed as representing the official views or policies of IARPA, the ODNI, or the U.S. Government.

This material was funded in part by the U.S. Department of Energy, Office of Science, Office of Advanced Scientific Computing Research Quantum Testbed Pathfinder Program.

Sandia National Laboratories is a multimission laboratory managed and operated by National Technology \& Engineering Solutions of Sandia, LLC, a wholly owned subsidiary of Honeywell International Inc., for the U.S. Department of Energy’s National Nuclear Security Administration under contract DE-NA0003525. This paper describes objective technical results and analysis. Any subjective views or opinions that might be expressed in the paper do not necessarily represent the views of the U.S. Department of Energy or the United States Government.

\bibliographystyle{IEEEtran}
\bibliography{reference,referencesGSTQuantum}

\appendix
\subsection{Column Subset Selection Heuristics}
\label{app:cssp_heuristics}

As discussed in Section \ref{sec:pgg_FPR}, the first stage of the per-germ global FPR algorithm wherein we select for each germ a subset of its amplified parameters to require sensitivity to can be reduced to a problem known as the column subset selection problem (CSSP). CSSP arises naturally in many different domains, ranging from machine learning to (as is relevant in this work) optimal experiment design \cite{ordozgoiti2019regularized,de2011note,avron2013faster}. As such a significant amount of work has been performed studying heuristics for solving the CSSP. While reviewing that (quite large) body of research is beyond the scope of this paper, in this appendix we'll provide additional details on the specific heuristics we've used in this work. In Appendix \ref{app:greedy_search} we will discuss our implementation of the greedy search heuristic, and in Appendix \ref{app:rrqr} we'll discuss a heuristic based on the rank-revealing QR decomposition. In both cases the algorithms begin by taking as input the matrix $J$ described in Equation \ref{eqn:sing_vector_concat_matrix} and Algorithm \ref{alg:pgg_fpr_stage_1}.

\subsubsection{Greedy Search}
\label{app:greedy_search}

Given as input the matrix of concatenated right singular vectors for each germ's twirled derivative,

\begin{equation}
    J = \begin{bmatrix}V_{g_0} & V_{g_1} & \cdots & V_{g_{N_g}} \end{bmatrix},   
    \label{eqn:app_sing_vector_concat_matrix}
\end{equation}

\noindent our goal is to select a submatrix of $J$, $J_{\text{min}}$, formed from its columns such that $J_{\text{min}}$ has a column-rank equal to the number of amplifiable parameters and is well-conditioned. For the greedy search heuristic we do so by mapping the problem to a combinatorial optimization problem which we approximate the solution to using greedy search. For the greedy search heuristic we begin with a set of candidate vectors $\{\vec{v}_J\}$, i.e. the set of column vectors for $J$, and repeatedly ask the question: \textit{which vector $\vec{v}_{\text{cand}} \in \{\vec{v}_{J} \}$ improves my cost-function the most?} Here we choose a multi-objective cost function. Given either a complete or partial candidate solution set, $\{\vec{v}_{\text{cand}}\}$, the first objective function is the column-rank of $J_{\text{cand}}$, the matrix formed by concatenating the elements of $\{\vec{v}_{\text{cand}}\}$ column-wise. The second objective function is the trace of the pseudo-inverse of $J_{\text{cand}} J_{\text{cand}}^\top$,

\begin{equation}
    \text{Tr}\left( \left(J_{\text{cand}} J_{\text{cand}}^\top\right)^+ \right),
    \label{eqn:pinv_trace}
\end{equation}

\noindent where superscript $+$ denotes the Moore-Penrose pseudo-inverse, and is used to ensure the well-conditionedness of our solution. This cost-function arises naturally in statistics in optimal experiment design theory where the condition is called A-optimality \cite{pukelsheim2006optimal}. When comparing two candidate solutions using these two objective functions the first objective function (i.e. the rank) is given preference, with the latter used to distinguish among candidate solutions of the same rank.

While conceptually simple, the greedy search algorithm if implemented naively as described is far too computationally expensive to be useful for even a modestly-sized two-qubit gate set, as repeatedly calculating the rank and pseudo-inverse for every possible new update to the candidate solution set quickly becomes intractable. The search can be sped up significantly by leveraging the fact that the evaluation of the cost function corresponding to the addition of a single vector to a candidate solution set corresponds to a rank-one update. Let $\vec{v}_{\text{up}}$ denote a candidate addition to the solution set and let $(J_{\text{cand}} J_{\text{cand}}^\top)^+$ denote the pseudo-inverse of the current candidate solution matrix's gramian. The rank-one update to the pseudo-inverse is given by \cite{meyer1973generalized} 

\begin{equation}
    \left(J_{\text{cand}} J_{\text{cand}}^\top + \vec{v}_{\text{up}} \vec{v}_{\text{up}}^\top \right)^+ = \left(J_{\text{cand}} J_{\text{cand}}^\top\right)^+ + \Gamma,
\end{equation}

\noindent where the value for $\Gamma$ has two different cases\footnote{For symmetric updates there are technically three cases, but only two of these are relevant here. See \cite{meyer1973generalized}.}:


\begin{equation}
    \Gamma = 
    \begin{cases}
        \frac{-1}{\| \vec{y} \|^2} \left( \vec{x} \vec{y}^\top + \vec{y} \vec{x}^\top  \right) + \frac{\beta}{\|\vec{y}\|^4} \vec{y}\vec{y}^\top & \|\vec{y}\| \neq 0, \\
        \frac{-\beta}{|\beta|^2} \vec{x} \vec{x}^\top & \|\vec{y}\|=0 ; \beta\neq 0,
    \end{cases}
    \label{eqn:pinv_update_rules}
\end{equation}

\noindent where

\begin{equation*}
    \begin{aligned}
    \vec{x} &= \left(J_{\text{cand}} J_{\text{cand}}^\top\right)^+ \vec{v}_{\text{up}}, \\
    \vec{y} &= \left(I - \left(J_{\text{cand}} J_{\text{cand}}^\top\right) \left(J_{\text{cand}} J_{\text{cand}}^\top\right)^+ \right) \vec{v}_{\text{up}},\\
    \beta &= 1+ \vec{v}_{\text{up}}^\top  \left(J_{\text{cand}} J_{\text{cand}}^\top\right)^+ \vec{v}_{\text{up}}.
    \end{aligned}
\end{equation*}

\noindent Usefully, the rank-one pseudo-inverse update also gives us the updated rank for free, as $\vec{y}$ as defined above corresponds to the projection of $\vec{v}_{\text{up}}$ onto the orthogonal complement of the column space of $\left(J_{\text{cand}} J_{\text{cand}}^\top\right)$, meaning that when $\|\vec{y}\| \neq 0$ this indicates that the rank of the updated matrix increases by one.

Using rank-one updates we can very efficiently evaluate each of the candidate additions to our solution set during each greedy search iteration, making the greedy search approach to column subset selection practical here. The greedy search algorithm is summarized in Algorithm \ref{alg:greedy_search_heuristic}.

In applying the greedy search algorithm to the problem of selecting reduced sets of fiducial pairs as described in Algorithm \ref{alg:pgg_fpr_stage_2}, we proceed in largely the same manner as described in Algorithm \ref{alg:greedy_search_heuristic}, but since our low-rank updates are no longer generally rank-one, we instead need to use a more general set of pseudo-inverse update rules due to Minamide \cite{minamide1985eension}.

\begin{algorithm}
\caption{Greedy Search Algorithm}\label{alg:greedy_search_heuristic}
\SetKwInOut{Input}{Input}\SetKwInOut{Output}{Output}
\Input{Matrix $J$ from Eqn. \ref{eqn:app_sing_vector_concat_matrix} of concatenated right singular vectors of each germ's twirled derivative.}
\Output{Matrix $J_{\text{min}}$, a submatrix of $J$ formed from its columns that is full column rank and well-conditioned.}

$\{\vec{v}_{\text{cand}} \} \gets \{\}$ \Comment*{Candidate solution set initially empty}
$J_{\text{cand}} \gets  [\hspace{.25em}]$\;
$\text{bestscore} \gets 0$\;
\While{\texttt{rank}$(J_{\text{cand}}) <$  \texttt{rank}$(J) \vee J_{\text{cand}}=[\hspace{.25em}]$}{
    $(J_{\text{cand}} J_{\text{cand}}^\top)^+ \gets$ \texttt{pinv}$(J_{\text{cand}} J_{\text{cand}}^\top)$ \Comment*{Skip on first iteration when $J_{\text{cand}}$ is empty}
    \For{$\vec{v}_{\text{up}} \in (\{\vec{v}_J\} - \{\vec{v}_{\text{cand}}\})$}{
        $\text{newscore, newrank} \gets $ \texttt{pinv\_trace\_update}$((J_{\text{cand}} J_{\text{cand}}^\top)^+ , \vec{v}_{\text{up}} )$    \Comment*{Eqns. \ref{eqn:pinv_trace}--\ref{eqn:pinv_update_rules}} 
        \If{($\text{newrank}>$ \texttt{rank}$(J_{\text{cand}})) \wedge (\text{newscore}>\text{bestscore})$}{
            $\text{bestscore} \gets \text{newscore}$\;
            $\vec{v}_{\text{best}} \gets \vec{v}_{\text{up}}$ \Comment*{Best found addition}
        }
    }
    $\{\vec{v}_{\text{cand}}\} \gets \vec{v}_{\text{best}}$\;
    $J_{\text{cand}} \gets$ \texttt{concatenate}$(J_{\text{cand}}, \vec{v}_{\text{best}})$\;
}
$J_{\text{min}} \gets J_{\text{cand}}$
\end{algorithm}

\subsubsection{Rank-Revealing QR Decompositions}
\label{app:rrqr}

Another class of heuristics that has seen significant development for column subset selection is based on the application of the rank-revealing QR decomposition \cite{chan1992some}. An advantage of this class of protocols is that (at least for the simpler variants) they can be phrased entirely linear-algebraically, and as such allow us to straightforwardly leverage high-performance numerical linear algebra packages to implement them efficiently. We have found that this additional computational efficiency has allowed us in practice to scale up the first stage of the per-germ global FPR algorithm beyond what is practical to do with greedy search (even with low-rank update techniques) to experiment designs for three-qubit and two-qutrit GST.

For this work we use a heuristic due to Golub, Klema and Stewart (GKS) \cite{golub1976rank, golub1996matrix}, which we briefly summarize. The following theorem forms the basis of the GKS algorithm:

\begin{theorem}[Thm. 12.2.1 \cite{golub1996matrix}]
Let $A$ be an $m \times n$ real-valued matrix, with an SVD given by $A=U \Sigma V^\top$, and rank $\tilde{r}$. Define the matrix $B_1 \in \mathbb{R}^{m \times \tilde{r}}$ as
\begin{equation*}
    AP = \begin{bmatrix} B_1 &  B_2\end{bmatrix},
\end{equation*}

\noindent where $P \in \mathbb{R}^{n \times n}$ is a column permutation. If

\begin{equation*}
    P^\top V = \begin{bmatrix} \tilde{V}_{11} & \tilde{V}_{12}\\ \tilde{V}_{21} & \tilde{V}_{22}  \end{bmatrix}
\end{equation*}

\noindent and $\tilde{V}_{11}$ is nonsingular then

\begin{equation*}
    \frac{\sigma_{\tilde{r}} (A) }{\| \tilde{V}_{11}^{-1} \|_2 } \leq \sigma_{\tilde{r}} (B_1) \leq \sigma_{\tilde{r}} (A),
\end{equation*}

\noindent where $\|\cdot\|_2$ denotes the spectral norm and $\sigma_{\tilde{r}}$ denotes the $\tilde{r}$\textsuperscript{th} singular value of the specified matrix.

\label{thm:golub_svd}
\end{theorem}

From Theorem \ref{thm:golub_svd} it follows that if our goal is to identify a subset of columns from $A$ such that $B_1$ is well-conditioned and full rank we can do so by identifying a column permutation $P$ that when applied $A$'s right singular vectors gives a well-conditioned $\tilde{r} \times \tilde{r}$ sub-block $\tilde{V}_{11}$.

The last observation needed to turn Theorem \ref{thm:golub_svd} into a full algorithm is that identifying a column permutation with the aforementioned properties is exactly what the heuristics used in the rank-revealing QR decomposition are designed for. More concretely, given a matrix $M$ the rank-revealing QR decomposition of $M$ generates a set of three matrices, $Q$, $R$ and $P$ such that

\begin{equation*}
    M P = Q R,
\end{equation*}

\noindent where $P$ is a column permutation matrix, $Q$ is orthogonal and $R$ is an upper triangular matrix with strictly non-increasing diagonal entries and where the number of non-zero diagonal entries is equal to $\text{rank}(M)$. Applying the rank-revealing QR decomposition to the matrix $ \begin{bmatrix} V_{11}^\top & V_{21}^\top \end{bmatrix}$ yields

\begin{equation}
    \begin{bmatrix} V_{11}^\top & V_{21}^\top \end{bmatrix} P = Q \begin{bmatrix} R_{11}  & R_{12} \end{bmatrix}.
    \label{eqn:rrqr}
\end{equation}

\noindent Finally, following from the norm preserving properties of the QR decomposition it follows that $\| V_{11}^\top \| = \|R_{11}\|$, and so insofar as the permutation constructed by the rank-revealing QR decomposition is successful in constructing a well-conditioned $R_{11}$ matrix it follows that this permutation likewise results in a well-conditioned $V_{11}$ subblock, and ultimately in a well-conditioned subset of column vectors from $A$. The details of this algorithm are summarized in Algorithm \ref{alg:rrqr_cssp}.

\begin{algorithm}
\caption{Rank-Revealing QR Algorithm for CSSP}\label{alg:rrqr_cssp}
\SetKwInOut{Input}{Input}\SetKwInOut{Output}{Output}
\Input{Matrix $J$ from Eqn. \ref{eqn:app_sing_vector_concat_matrix} of concatenated right singular vectors of each germ's twirled derivative.}
\Output{Matrix $J_{\text{min}}$, a submatrix of $J$ formed from its columns that is full column rank and well-conditioned.}

$U, \Sigma, V^\top \gets$ \texttt{SVD}($J$)\;
$\tilde{r} \gets$ \texttt{rank}$(J)$\;
$V' \gets V^\top[0:\tilde{r}, :]$ \Comment*{Take the first $\tilde{r}$ rows of $V^\top$}
$P, Q, R \gets $ \texttt{rank\_revealing\_QR}$(V')$ \Comment*{Eqn. \ref{eqn:rrqr}}
$J_{\text{min}}\gets (JP)[:, 0:\tilde{r}]$ \Comment*{Take first $\tilde{r}$ columns of the permuted $J$ matrix for $J_{\text{min}}$}

\end{algorithm}

\end{document}